\documentclass[aps,pra,floatfix,twocolumn,nofootinbib,showpacs]{revtex4}
\usepackage{graphicx} \usepackage{amsmath} \usepackage{amssymb}

\newcommand{\BEQ}{\begin{eqnarray}}
\newcommand{\EEQ}{\end{eqnarray}}
\newcommand{\BEA}{\begin{eqnarray}}
\newcommand{\EEA}{\end{eqnarray}}

\renewcommand{\d}{{\rm d}}

\newcommand{\eps}{\varepsilon}

\newcommand{\tr}{{\rm tr}}

\renewcommand{\S}{{\rm \bf C}}
\newcommand{\R}{{\rm \bf H}}

\newcommand{\Q}{{\cal Q}}

\newcommand{\diag}{{\rm diag}}
\newcommand{\ssum}{{\sum}}

\newcommand{\comment}[1]{}
\newcommand{\lb}{\left[}
\newcommand{\lp}{\left(}
\newcommand{\rb}{\right]}
\newcommand{\rp}{\right)}

\begin{document}

\draft
\title
{Thermodynamics of enhanced heat transfer: a model study}
\date{\today}
\author{ Karen Hovhannisyan and Armen E. Allahverdyan}
\affiliation{
Yerevan Physics Institute, Alikhanian Brothers Street 2, Yerevan 375036, Armenia}

\begin{abstract} Situations where a spontaneous process of energy or
matter transfer is enhanced by an external device are widespread in
nature (human sweating system, enzyme catalysis, facilitated diffusion
across bio-membranes, industrial heat exchangers). The thermodynamics of
such processes remains however open. Here we study enhanced heat
transfer by a model junction immersed between two thermal baths at
different temperatures $T_h$ and $T_c$ ($T_h>T_c$).  The transferred
heat power is enhanced via controlling the junction by means of external
time-dependent fields.  Provided that the spontaneous heat flow process
is optimized over the junction Hamiltonian, any enhancement of this
spontaneous process does demand consumption and subsequent dissipation
of work. The efficiency of enhancement is defined via the increment
in the heat power divided over the amount of consumed work. We show that
this efficiency is bounded from above by $T_c/(T_h-T_c)$. Formally this
is identical to the Carnot bound for the efficiency of ordinary
refrigerators which transfer heat from cold to hot. It also shares some
(but not all) physical features of the Carnot bound. 

\end{abstract}


\comment{
Energy conversion, 84.60.-h

Irreversible thermodynamics, 05.70.Ln

Thermodynamics, 05.70.-a

Heat engines, 07.20.Pe

Refrigeration, 07.20.Mc

quantum statistical mechanics, 05.30.-d

}

\maketitle

\section{Introduction}

Heat spontaneously flows from hot to cold. In some cases it is necessary
to reverse this flow. Devices that perform this operation need an
external input of high-graded energy (work), which is lost
in the process: refrigerators cool a colder
body in the presence of a hotter environment, while heaters heat up a
hot body in the presence of a colder one \cite{callen}. The efficiency
(or coefficient of performance) of these devices is naturally defined as
the useful effect|for refrigerators this is the heat extracted from the
colder body, while for heaters this is the heat delivered to the hotter
body|divided over the work consumed per cycle from the work-source
\cite{callen}. The first and second laws of thermodynamics limit this
efficiency from above by the Carnot value: For a refrigerator (heater)
operating between two thermal baths at temperatures $T_c$ and $T_h$,
respectively, the Carnot efficiency reads \cite{callen}
\BEA
\zeta_{\rm refrigerator}= \frac{\theta}{1-\theta}, ~~~ \zeta_{\rm heater}= \frac{1}{1-\theta},
~~~ \theta\equiv \frac{T_c}{T_h}<1.
\label{arch}
\EEA

There are however situations, where the spontaneous direction of the
process is the desired one, but its power has to be increased. An
example of such a process is perspiration (sweating) of mammals
\cite{guyton}. A warm mammalian body placed in a colder environment will
naturally cool due to spontaneous heat transfer from the body surface.
Three spontaneous processes are involved in this: infrared
radiation, conduction and convection \cite{guyton}. When the
environmental temperature is not very much lower than the body
temperature, the spontaneous processes are not sufficiently powerful,
and the sweating mechanism is switched on: sweating glands produce
water, which during evaporation absorbs latent heat from the body
surface and thus cools it \cite{guyton}. Some amount of free energy (work) is spent in
sweating glands to wet the body surface. Similar examples of heat transfer
are found in the field of industrial heat-exchangers, where the
external source of work is employed for mixing up the heat-exchanging fluids. 

The main feature of these examples is that they combine spontaneous and
driven processes. Both are macroscopic, and with both of them the work
invested in enhancing the process is ultimately consumed and dissipated.
Pertinent examples of enhanced transport exist in biology
\cite{blum,kovac}. During enzyme catalysis, the spontaneous rate of a
chemical reaction is increased due to interaction of the corresponding
enzyme with the reaction substrate. (A chemical reaction can be regarded
as particle transfer from a higher chemical potential to a lower one.)
There are situations where enzyme catalysis is fueled by external
sources of free energy supplied by co-enzymes \cite{blum}. However, many
enzymes function autonomously and cyclically: The enzyme gathers free
energy from binding to the substrate, stores this free energy in slowly
relaxing conformational degrees of freedom \cite{mikhailov,vancelow},
and then employs it for lowering the activation barrier of the reacion
thereby increasing its rate \cite{blum,kovac,mikhailov,vancelow}.
Overally, no free energy (work) is consumed for enhancing the process
within this scenario. Similar situations are realized in transporting
hydrophilic substances across the cell membrane \cite{blum}. Since these
substances are not soluble in the membrane, their motion along the
(electro-chemical) potential gradient is slow, and special transport
proteins are employed to enhance it \cite{blum,kovac}. Such a {\it
facilitated diffusion} normally does not consume free energy (work). 

These examples of enhanced processes motivate us to ask several
questions. Why is that some processes of enhancement employ work
consumption, while others do not? When enhancement does (not) require
work consumption and dissipation? If the work-consumption does take
place, how to define the efficiency of enhancement, and are there bounds
for this efficiency comparable to (\ref{arch})? These questions belong
to thermodynamics of enhanced processes, and they are currently open.
Laws of thermodynamics do not answer to them directly, because here the
issue is in increasing the rate of a process. Dealing with time-scales
is a weak-point of the general thermodynamic reasoning
\cite{lindblad}, a fact that motivated the development of finite-time
thermodynamics \cite{ftt}. 

Here we address these questions via analysing a quantum model for
enhanced heat transfer. The model describes a few-level junction
immersed between two thermal baths at different temperatures; see
section \ref{model}. The junction is subjected to an external field,
which enhances the heat transferred by the junction along its
spontaneous direction.  The virtue of this model is that the
optimization of the transferred heat over the junction Hamiltonian can
be carried out explicitly. Based on this, we determine under which
conditions the enhancement of heat-transfer does require
work-consumption.  We also obtain an upper bound on the efficiency of
enhancement, which in several aspects is similar to the Carnot bound
(\ref{arch}). 

Heat flow in microscale and nano-scale junctions received much attention
recently
\cite{q_t,lepri,segal_nitzan,dhar,segal,ren_b_li,mahler,feldman,AKM}.
This is related to the general trend of technologies towards smaller
scales. Needless to stress that thermodynamics of enhanced heat-transfer
is relevant for this field, because it should ultimately draw the
boundary beetween what is possible and what is not when cooling a hot
body in the presence of a colder one. Brownian pumps is yet another
field, where external fields are used to drive transport; see, e.g.,
\cite{parrondo,no_pumping} and references therein. Some of the set-ups
studied in this field are not far from the enhanced heat transport
investigated here. However, thermodynamical quantities (such as work and
enhancement efficiency) were so far not studied for these systems,
though thermodynamics of Brownian motors [work-extracting devices] is a
developed subject reviewed in \cite{cisneros}.

The rest of this paper is organized as follows. The model of
heat-conducting junction is introduced in section \ref{model}. Section
\ref{QforE} shows how the transferred heat (with and without enhancing)
can be optimized over the junction structure. The efficiency of
enhancing is studied in section \ref{effi}. Section \ref{baratino}
discusses how some of the obtained results can be recovered from the
formalism of linear non-equilibrium thermodynamics. We summarize in
section \ref{conc}. Several questions are relegated to Appendices.

\section{The model}
\label{model}

\begin{figure}
\includegraphics[width=8.8cm]{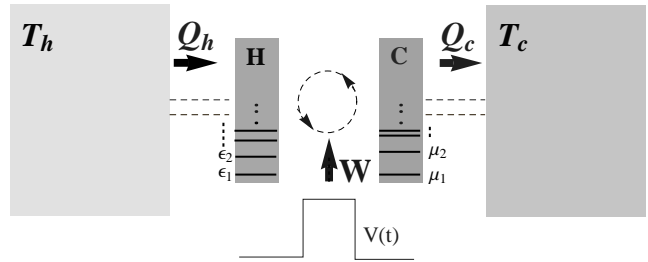}
\vspace{0.1cm}
\caption{ 
The heat pump model. The few-level systems $\R$ and $\S$ operate between
two baths at temperatures $T_c$ and $T_h$ $T_c<T_h$.  During the first
step of operation the two systems interact together either spontaneously
or driven by a work-source at the cost of work $W$. During this stage
couplings with the thermal baths is neglected (thermal isolation). In
the second step the systems $\R$ and $\S$ do not interact with each
other and freely relaxes to their equilibrium states (\ref{1}) under
action of the corresponding thermal bath. 
}
\label{f1}
\end{figure}

Our model for the heat pump (junction) consists of two quantum systems
$\R$ and $\S$ with Hamiltonians $H_\R$ and $H_\S$, respectively; see
Fig.~\ref{f1}. Each system has $n$ energy levels and couples to its
thermal bath. Similar models were employed for studying heat engines
\cite{jmod,armen} and refrigerators \cite{AKM}. It will be seen below
that this model admits a classical interpretation, because all the
involved initial and final density matrices will be diagonal in
the energy representation. We shall however work within the 
quantum framework, since it is more intuitive.

Initially, $\R$ and $\S$ do not interact with each other. Due to
coupling with their baths they are in equilibrium at temperatures
$T_h=1/\beta_h>T_c=1/\beta_c$ [we set $k_{\rm B}=1$]:
\BEA
\label{1}
\rho={e^{-\beta_h H_\R}}/{\tr\, [e^{-\beta_h H_\R}]}, ~~
\sigma={e^{-\beta_c H_\S}}/{\tr\, [e^{-\beta_c H_\S}]},
\EEA
where $\rho$ and $\sigma$ are the initial Gibbsian density matrices of $\R$ and $\S$, respectively.
We write
\begin{gather}
\label{2}
\rho=\diag [r_n,...,r_1], ~~~~~~~~~~
\sigma=\diag [s_n,...,s_1], \\
H_\R=\diag [\eps_n,...,\eps_1=0\,],~~
H_\S=\diag [\mu_n,...,\mu_1=0\,], \nonumber
\end{gather}
where $\diag[a,..,b]$ is a diagonal matrix with entries $(a,...,b)$, and
where without loss of generality we have nullified the lowest energy level
of both $\R$ and $\S$.  Thus the overall initial density matrix is
\BEA
\label{111}
\Omega_{\rm in}=\rho\otimes\sigma, 
\EEA
and the initial Hamiltonian of the junction is
\BEA
\label{222}
H_0=H_\R\otimes 1+1\otimes H_\S .
\EEA

\subsection{Spontaneous operation}
\label{sponto}

During a spontaneous process no work is exchanged with external sources.
For our situation a spontaneous heat transfer will amount to a certain
interaction between $\R$ and $\S$. Following to the approach of
\cite{partovi,mityugov,Dom} we model this interaction via a Hamiltonian
that conserves the (free) Hamiltonian $H_0$ [see (\ref{222})] for
all interaction times. This then realizes the main premise of
spontaneous processes: no work exchange at any time. Our model for
spontaneous heat transfer consists of two steps. 

{\bf 1.} During the first step $\R$ and $\S$ interact with each
other [collision]. We assume that this interaction takes a sufficiently short time
$\delta$, and during this time the coupling with the two
thermal baths can be neglected [thermal isolation]. The interaction is
described by the Hamiltonian $H_{\rm int}$ added to (\ref{222}):
\BEA
\label{202}
H=H_\R\otimes 1+1\otimes H_\S+H_{\rm int} .
\EEA
The overall Hamiltonian $H$ again lives in the $n^2$-dimensional Hilbert
space of the junction \footnote{ More precisely, we had to write the
Hamiltonian (\ref{202}) as $H_\R\otimes 1+1\otimes H_\S+\alpha(t)H_{\rm
int}$, where $\alpha(t)$ is a switching function that turns to zero both
at the initial and final time. This will however not alter the
subsequent discussion in any serious way. }. 

As argued above, the interaction Hamiltonian
commutes with the total Hamiltonian:
\BEA
[H_0,H_{\rm int}] =0,
\label{co}
\EEA
making the energy $H_0$ a conserved quantity \footnote{This
implementation of spontaneous heat-transfer processes admits an
obvious generalization: one need not require the conservation of $H_0$
for all interaction times, it suffices that no work is consumed or
released within the overall energy budget of the process in the
time-interval $[0,\delta]$. For our purposes this generalization will
not be essential; see (\ref{miller}).}.  To have a non-trivial effect on
the considered system, the interaction Hamiltonian $H_{\rm int}$ should
not commute with the separate Hamiltonian: $[H_\R\otimes 1,H_{\rm int}]
\not=0$. For this to be the case the spectrum of $H_0$ should contain at
least one degenerate eigenvalue. Otherwise, relations $[H_0,H_{\rm int}]
=0$ and $[H_\R\otimes 1,H_0] =0$ will imply $[H_\R\otimes 1,H_{\rm int}]
=0$ (and thus a trivial effect of $H_{\rm int}$), because the eigen-base
of $H_0$ will be unique (up to re-numbering of its elements and their
multiplication by phase factors).  The energy
\BEA
Q_h^{\rm [sp]}=\tr \left(H_\R\left[\rho-\,{\rm tr}_\S\left(e^{-\frac{i\delta}{\hbar}H_{\rm int}  }\,\Omega_{\rm in} \,
e^{\frac{i\delta}{\hbar}H_{\rm int}  }   \right)\,\right]\right),
\label{kora}
\EEA
lost by $\R$ during the interaction is gained by $\S$. Here ${\rm tr}_\R$ and ${\rm tr}_\S$ are the partial traces.

Commutative interaction Hamiltonians (\ref{co}) are applied to studying heat
transfer in \cite{partovi,mityugov,Dom}. 
Refs.~\cite{partovi,mityugov} are devoted to supporting the
thermodynamic knowledge via quantum Hamiltonian models. In
contrast, the approach of \cite{Dom} produced new results.

{\bf 2.} For times larger than $\delta$, $\R$ and $\S$ do not interact
and freely relax back to their equilibrium states (\ref{1}, \ref{111})
due to interaction with the corresponding thermal baths. These
equilibrium states are reached after some relaxation time $\tau$. Thus
the cycle is closed|the junction returns to its initial state|and
$Q_h^{\rm [sp]}$ given by (\ref{kora}) is the heat per cycle taken from
the hot thermal bath during the relaxation (and thus during the overall cycle). 

It should be obvious that once $T_h>T_c$ we get $Q_h^{\rm [sp]}>0$: heat
spontaneously flow from hot to cold. The proof of this fact is given in
\cite{partovi,mityugov,Dom,armen,AKM}. 

For times larger than $\tau$ there comes another interaction pulse
between $\R$ and $\S$, and the cycle is repeated. 

\subsubsection{Power}
\label{pow}

Recall that the power of heat-transfer is defined as the ratio of the
transferred heat to the cycle duration $\tau$, $Q_h^{\rm [sp]}/\tau$.
For the present model $\tau$ is mainly the duration of the second stage,
i.e., $\tau$ is the relaxation time, which depends on the concrete
physics of the system-bath coupling. For a weak system-bath coupling
$\tau$ is larger than the internal characteristic time of $\R$ and $\S$.
In contrast, for the collisional system-bath interaction, $\tau$ can be
very short; see Appendix \ref{po}. Thus the cycle time $\tau$ is finite,
and the power $Q_h^{\rm [sp]}/\tau$ does not vanish due to a large cycle
time, though it can vanish due to $Q_h^{\rm [sp]}\to 0$. 

Note that some entropy is produced during the free relaxation. This
entropy production can be expressed via quantities introduced in
(\ref{111}--\ref{kora}); see \cite{AKM} for details. The first step does
not produce entropy, because it is thermally isolated and is realized by
a unitary operation that can be reversed by operating only on observable
degrees of freedom ($\R+\S$). It is seen that no essential aspect of the
considered model depends on details of the system-bath interaction. This
is an advantage. 

\subsection{Driven operation}

The purpose of driving the junction with an external field is to enhance
[increase] the spontaneous heat $Q_h^{\rm [sp]}$. The driven
regime reduces to the above two steps, but instead of the spontaneous
interaction we have the following situation: the interaction between $\R$ and
$\S$ is induced by an external work-source. Thus (\ref{co}) does not
hold anymore.  The overall interaction [between $\R$, $\S$ and the
work-source] is described via a time-dependent potential $V(t)$ in the
total Hamiltonian
\BEA
\label{toto}
H_\R\otimes 1+1\otimes H_\S+V(t) 
\EEA
of $\R+\S$. The interaction process
is still thermally isolated: $V(t)$ is non-zero only in a short
time-window $0\leq t\leq \delta$ and is so large there that the
influence of the couplings to the baths can be neglected. 

Thus the dynamics of $\R+\S$ is unitary for $0\leq t\leq \delta$:
\BEA
\Omega_{\rm f}\equiv \Omega (\delta )={\cal U}\,
\Omega_{\rm i}\, {\cal U}^\dagger, ~~{\cal U}={\cal T}e^{-\frac{i}{\hbar} \int_0^\delta \d s \lb V(s)+H_0\rb  },
\label{barbos}
\EEA
where $\Omega_{\rm i}=\Omega(0)=\rho\otimes\sigma$ is the initial state defined in
(\ref{1}), $\Omega_{\rm f}$ is the final density matrix, ${\cal
U}$ is the unitary evolution operator, and where ${\cal T}$ is the time-ordering operator.
The work put into $\R+\S$ reads \cite{lindblad,callen}
\BEA
W=E_{\rm f}-E_{\rm i}=\tr [\,( H_\R\otimes 1+1\otimes H_\S)\, (\Omega_{\rm f}-\Omega_{\rm i}) \,],
\label{work}
\EEA
where $E_{\rm f}$ and $E_{\rm i}$ are initial and final energies of $\R+\S$.
Due to the interaction, $\R$ ($\S$) looses (gains) an amount of energy $Q_h$ ($Q_c$)
\BEA
\label{heats}
Q_h=\tr (\,H_\R[\,\rho-\,{\rm tr}_\S\,\Omega_{\rm f}\,]\,), \\
Q_c=\tr (\,H_\S\,[{\rm tr}_\R\,\Omega_{\rm f}-\sigma\,]\,).
\label{frid}
\EEA
Eqs.~(\ref{work}, \ref{heats}) imply an obvious relation
\BEA
W=Q_c-Q_h.
\EEA

Recall that for spontaneous processes not only the consumed work is zero, $W=0$, but 
also the stronger conservation condition (\ref{co}) holds.

Once $\R+\S$ arrives at the final state $\Omega_{\rm fin}$, the
inter-system interaction is switched off, and $\R$ and $\S$ separately
[and freely] relax to the equilibrium states (\ref{1}). During this
process $Q_h$ is taken as heat from the hot bath, while $Q_c$ is given
to the cold bath. Note from section \ref{pow} that the driven
operation does not change the cycle time $\tau$, because the latter
basically coincides with the relaxation time. Therefore, increasing
$Q_h$ means increasing heat transfer power. 

\section{Maximization of heat}
\label{QforE}

\subsection{Maximization variables}

The type of questions asked by thermodynamics concerns limiting, optimal
characteristics. Sometimes the answers are uncovered directly via the basic
laws of thermodynamics, an example being the Carnot bound (\ref{arch}).
However, more frequently than not, this goal demands explicit
optimization procedures \cite{ftt}. 

We shall maximize the heat $Q_h$ transferred from the hot bath over the
full Hamiltonian of the junction. For spontaneous
processes this amounts to maximizing over Hamiltonian (\ref{202}) living
in the $n^2$-dimensional Hilbert space of the junction $\R+\S$ and satisfying
condition (\ref{co}). For driven processes we shall maximize over
Hamiltonians (\ref{toto}). In this case we shall impose an additional
condition that the work put into $\R+\S$ in the first step does not
exceed $E>0$:
\BEA
\label{si}
W\leq E.
\EEA
Once the work put into the system is a resource, it is natural to
operate with resources fixed from above. 

Recall that the Hamiltonians (\ref{202}, \ref{toto}) live in the
$n^2$-dimensional Hilbert space.  The bath temperatures $T_c$ and $T_h$
and the dimension $n^2$ (the number of energy levels) will be held fixed
during the maximization. 

Due to (\ref{kora}) the maximization of the spontaneous heat $Q_h^{\rm
[sp]}$ over the Hamiltonians (\ref{202}, \ref{co}) amounts to maximizing
over unitary operators $e^{\frac{i\delta}{\hbar}H_{\rm int} }$, and over
the energies $\{\eps_k\}_{k=2}^n$, $\{\mu_k\}_{k=2}^n$ of $\R$
and $\S$. Likewise, as seen from (\ref{toto}--\ref{work}), the
maximization of the driven heat $Q_h$ amounts to maximizing under
condition (\ref{si}) over all unitary operators ${\cal U}$ living in the
$n^2$-dimensional Hilbert space, and over the energies
$\{\eps_k\}_{k=2}^n$, $\{\mu_k\}_{k=2}^n$.

We should stress that the class of Hamiltonians living in the
$n^2$-dimensional Hilbert space [with or without constraint (\ref{co})] is
well-defined due to separating the heat transfer into two steps
(thermally isolated interaction and isothermal relaxation). More general
classes of processes can be envisaged. For instance, we may write the
free Hamiltonian as $H_0+H_{B,c}+H_{B,h}$, where $H_0$, $H_{B,c}$ and
$H_{B,h}$ are, respectively, the Hamiltonians of the junction and the
two thermal baths. Now the Hamiltonian $H_{\rm int}$ of spontaneous
processes will be conditioned as $[H_{\rm int}, H_0+H_{B,c}+H_{B,h}]=0$.
This condition is more general than (\ref{co}). Then the corresponding
class of driven processes can be naturally defined via the same class of
Hamiltonians but without this commutation condition. We do not consider
here such general processes, since we are not able to optimize them. 

\subsection{Unconstrainted maximization}

As seen below, the maximization of the spontaneously transferred heat
(\ref{kora}) amounts to a particular case of maximizing $Q_h$. So we
shall directly proceed to maximizing the driven heat $Q_h$; see
(\ref{heats}). 

First, take in (\ref{si}) the simplest case: $E=+\infty$. This case contains the
pattern of the general solution.  Here we have no constraint on
maximization of $Q_h$ and it is done as follows. Since $\tr [H_\R\rho]$
depends only on $\{\eps_k\}_{k=2}^n$, we choose $\{\mu_k\}_{k=2}^n$ and
$V(t)$ so that the final energy $\tr [H_\R\,{\rm tr}_\S\Omega_{\rm f}]$
attains its minimal value zero.  Then we shall maximize $\tr [H_\R\rho]$
over $\{\eps_k\}_{k=2}^n$. Note from (\ref{2})
\BEA
H_\R\otimes 1 &=& {\rm diag}[\,\eps_1\,\,,\ldots,\,\,\,\eps_1,\ldots,
                             \,\,\eps_n\,\,\,,\ldots,\,\,\eps_n\,\, ],\nonumber\\
\Omega_{\rm i}=\rho\otimes\sigma &=& {\rm
  diag}[\, r_1s_1,\ldots,r_1s_n,\ldots,
                            r_n s_1,\ldots,r_n s_n\, ].\nonumber
\EEA
It is clear that ${\rm tr} \left[ H_\R \, {\rm tr}_\S \Omega_{\rm f} \right] = 
{\rm tr} \lb ( H_\R \otimes 1 ){\cal U}\, \Omega_{\rm i} \, {\cal U}^\dagger\, \right]$ goes to zero when, e.g., $s_2=\ldots=s_n\to 0$
($\mu\equiv\mu_2=\ldots=\mu_n\to \infty$),
while ${\cal U}$ amounts to the SWAP operation
${\cal U}\rho\otimes\sigma{\cal U}^\dagger = \sigma\otimes\rho$.
Simple symmetry considerations show that at the maximum of the initial energy
$\tr [H_\R\sigma]$ the second level is $n-1$ fold degenerate, i.e. $\eps\equiv\eps_2=\ldots=\eps_n$. 
Denoting
\BEA
\label{burundi}
u= e^{-\beta_h \eps} \propto r_2=..=r_n
\EEA
we obtain for $Q_h={\cal Q}_h(\infty)$
\BEA
\label{10}
{\cal Q}_h(\infty) = T_h \ln\lp\frac{1}{u}\rp\lb1-\frac{1}{1+(n-1)u}\rb
\EEA
where $u$ is to be found from maximizing the RHS of (\ref{10}) over $u$, i.e., $u$ is
determined via
\BEA
1+(n-1)u+\ln u=0.
\label{capo}
\EEA
Note that in this case $W=+\infty$. 
In the $n\gg1$ limit we have 
$u=\frac{\ln n}{n}\lb 1+o(1)\rb$ from (\ref{capo}) and $Q_h=T_h\ln n\lb1+{\cal O}\lp\frac{\ln\ln n}{\ln n}\rp\rb$.

\subsection{Constrainted maximization}

The case of a finite $E$ in (\ref{si}) is more complicated.  We had to resort to
numerical recipes of Mathematica 6. Denoting $\{| i_\R\rangle
\}_{k=1}^n$ and $\{| i_\S\rangle \}_{k=1}^n$ for the eigenvectors of
$H_\R$ and $H_\S$, respectively, we see from (\ref{work}, \ref{heats})
that $W$ and $Q_h$ feel ${\cal U}$ only via the matrix
\BEA
\label{gomel}
C_{ij\,|\,kl}=|\langle i_\R j_\S |{\cal U}| k_\R l_\S\rangle|^2.
\EEA
This matrix is double-stochastic \cite{olkin}:
\BEA
\label{chernigov}
{\ssum}_{ij }C_{ij\,|\,kl}={\ssum}_{kl }C_{ij\,|\,kl}=1.
\EEA
Conversely, for {\it any} double-stochastic matrix $C_{ij\,|\,kl}$ there
is {\it some} unitary matrix $U$ with matrix elements $U_{ij\,|\,kl}$,
so that $C_{ij\,|\,kl}=|U_{ij\,|\,kl}|^2$ \cite{olkin}. Thus, when
maximizing various functions of $W$ and $Q_c$ over the unitary ${\cal
U}$, we can directly maximize over the $(n^2-1)^2$ independent elements
of $n^2\times n^2$ double stochastic matrix $C_{ij\,|\,kl}$. This fact
simplifies the problem. 

\begin{figure}
\includegraphics[width=8.5cm]{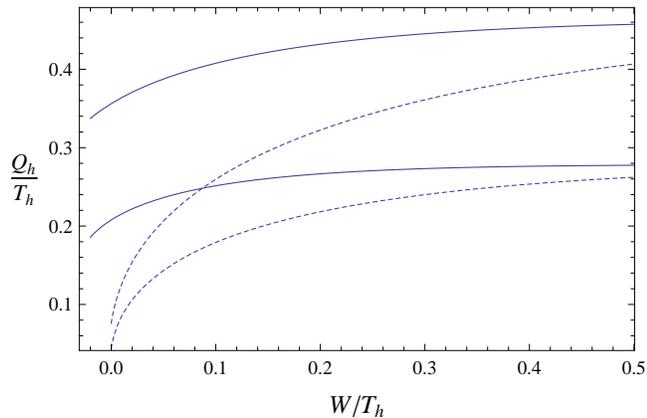}
\caption{ The optimal transferred heat ${\cal Q}_h$ versus work $W$. 
Dashed curves refer to $\theta\equiv T_c/T_h=0.9$: $n=2$ (lower 
dashed curve) and $n=3$ (upper dashed curve). Normal curves refer to $\theta=0.5$:
$n=2$ (lower normal curve) and $n=3$ (upper normal curve). }
\label{f22}
\end{figure}

Numerical maximization of $Q_h$ over all unitaries ${\cal
U}$|alternatively, over all doubly stochastic $C$ matrices
(\ref{gomel})|and energy spectra $\{\mu_k\}_{k=2}^{n}$ and
$\{\eps_k\}_{k=2}^{n}$ constrained by $W\leq E$ produced the following
results:

\begin{itemize}

\item The upper energy levels for both systems $\R$ and $\S$ are $n-1$ times degenerate [see (\ref{2})]:
\BEA
\label{petan}
\mu=\mu_2=\ldots=\mu_n,\quad \varepsilon=\varepsilon_2=\ldots=\varepsilon_n.
\EEA

\item The optimal unitary corresponds to SWAP: 
\BEA
{\cal U}\rho\otimes\sigma{\cal U}^\dagger = \sigma\otimes\rho.
\EEA

\item The work resource is exploited fully, i.e., the maximal $Q_h$ is reached for $W=E$.

\end{itemize}

Though we have numerically checked these results for $n\leq 5$ only, we
trust that they hold for an arbitrary $n$.

Denoting by $\Q_h$ the maximal value of $Q_h$, and introducing from (\ref{petan})
\BEA
\label{pelikan}
v=e^{-\beta_c\mu}\quad  {\rm and}\quad u=e^{-\beta_h\varepsilon},
\EEA
we have
\BEA
\label{tver}
&&\frac{\Q_h}{T_h}=\ln\lb\frac{1}{u}\rb\,\frac{(n-1)(u-v)}{\lb 1+(n-1)v \rb\lb 1+(n-1)u \rb}, \\ 
\label{tver1}
&&\frac{W}{T_h}=\frac{(\, \ln u - \theta\ln v\,)\,(n-1)(u-v)}{\lb 1+(n-1)v \rb\lb 1+(n-1)u \rb},
\EEA
where $u$ and $v$ in (\ref{tver}, \ref{tver1}) are determined from maximizing the RHS of (\ref{tver})
and satisfying constraint (\ref{tver1}). 

An important implication of (\ref{tver}, \ref{tver1}) is that $\Q_h(W)$ 
is an increasing function of $W$ for all allowed values of
$W$: 
\BEA
\label{elvonc}
\Q_h(W)>\Q_h(W')\quad {\rm if}\quad W>W'.
\EEA
Fig.~\ref{f22} illustrates this fact.  For fixed parameters $T_c$, $T_h$
and $n$, the allowed $W$'s range from a certain negative value|which
corresponds to work-extraction from a two-temperature system $\R+\S$|to
arbitrary $W>0$. Eq.~(\ref{elvonc}) expresses an intuitively expected,
but still non-trivial fact that the best transfer of heat takes place
upon consuming most of the available work. Note that this result holds
only for properly optimized values of ${\cal Q}_h$. One can find
non-optimal set-ups, where (\ref{elvonc}) is not valid \footnote{The
simplest example is a junction, where the free
Hamiltonian $H_0$ has a non-degenerate energy spectrum, and thus the
condition (\ref{co}) does not hold. There are no proper spontaneous
processes for this case. Still there can exist a work-exracting ($W<0$)
driven processes that transfer heat from hot to cold.}. 

\subsection{Optimization of spontaneous processes}

According to our discussion in section \ref{sponto}, the maximization of
transferred heat $Q_h^{\rm [sp]}$ given by (\ref{kora}) should proceed
over all unitary operators $e^{-\frac{i\delta}{\hbar}H_{\rm int} }$ with
$H_{\rm int}$ satisfying (\ref{co}) and over the energies
$\{\eps_k\}_{k=2}^n$, $\{\mu_k\}_{k=2}^n$ of $\R$ and $\S$.  This
maximization has been carried out along the lines described around
(\ref{chernigov}).  We obtained that the maximal spontaneous heat ${\cal
Q}_h^{\rm [sp]}$ is equal to ${\cal Q}_h$ in (\ref{tver}) under
condition $W=0$:
\BEA
\label{miller}
{\cal Q}_h^{\rm [sp]}={\cal Q}_h(W=0).
\EEA
Thus the optimal spontaneous processes coincide with optimal processes
with zero consumed work. This result is non-trivial, because the class
of unitary operators with $W=0$ is larger than the class of unitary
operators $e^{-\frac{i\delta}{\hbar}H_{\rm int} }$ with $H_{\rm int}$
satisfying (\ref{co}). Nevertheless, these two classes produce the 
same maximal heat.

\begin{itemize}

\item Eqs.~(\ref{elvonc}, \ref{miller}) imply that if the spontaneous
heat transfer process is already optimal (with respect to the junction
Hamiltonian) its enhancement with help of driven processes does demand
work-consumption, $W>0$.  This fact is non-trivial, because|as well known from
the theory of heat-engines|also work-extraction does lead to the heat
flowing from cold to hot \cite{callen,lindblad}.

\end{itemize}

Taking $W=0$ in (\ref{tver}, \ref{tver1}) and recalling (\ref{pelikan}) we get 
\BEA
\label{cond}
\mu=\varepsilon, \quad\quad 
u = v^{\theta}.
\EEA
The interpretation of (\ref{cond}) is that since there are only two
independent energy gaps in the system, they have to be precisely matched
for the spontaneous processes to be possible; see in this context the
discussion after (\ref{co}). Thus the spontaneous heat
${\cal Q}_h^{\rm [sp]}$ is given as 
\BEA
\label{qsp}
\frac{{\cal Q}_h^{\rm [sp]}}{T_c} = \ln\lb \frac{1}{v_0}\rb\frac{(n-1)(v_0^{\theta}-v_0)}{[1+(n-1)v_0^{\theta}][1+(n-1)v_0]},
\EEA
where $v_0$ maximizes the RHS of (\ref{qsp}).

\subsection{How much one can enhance the spontaneous process?}

We would like to compare the optimal spontaneous heat (\ref{qsp}) with
the optimal heat ${\cal Q}_{h}(\infty)$ transferred under consumption of
a large amount of work; see (\ref{10}, \ref{capo}) and recall
(\ref{elvonc}). One notes that for parameters of Fig.~\ref{f22} the
approximate equality ${\cal Q}_{h}(\infty)\approx {\cal Q}_{h}(W)$ is
reached already for $W/T_h<1$. This figure also shows that for the
temperature ratio $\theta\equiv T_c/T_h$ far from $1$, the improvement
of the transferred heat introduced by driving is not substantial. It is
however rather sizable for $\theta\simeq 1$, because here the
spontaneous heat (\ref{qsp}) is close to zero, while the heat ${\cal
Q}_{h}(\infty)$ does not depend on the temperature difference at all;
see Fig.~\ref{f22} and (\ref{10}, \ref{capo}). 

\section{Efficiency}
\label{effi}

\begin{figure}
\includegraphics[width=8.5cm]{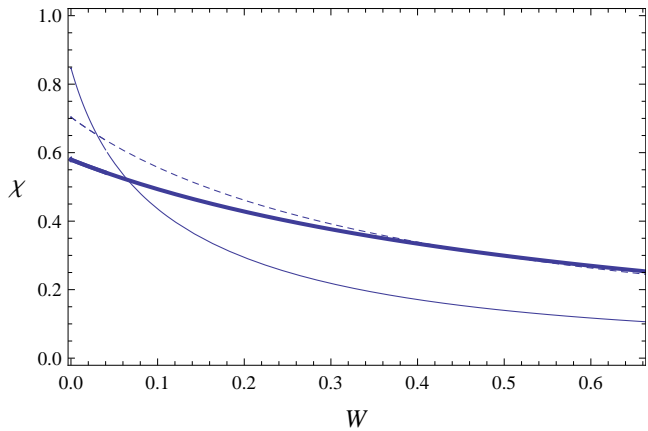}
\caption{ The efficiency $\chi$ versus work $W$ for $\theta\equiv T_c/T_h=0.5$ and $n=2$ (normal curve),
$n=10$ (dashed curve) and $n=30$ (thick curve).  }
\label{f2}
\end{figure}

We saw above that enhancing optimal spontaneous processes does
require work. Once this is understood, we can ask how efficient is this
work consumption. The efficiency is defined as
\BEA
\label{effa}
\chi(W)=\frac{\Q_h(W)-\Q_h^{\rm [sp]}}{W}>0,
\EEA
where $\Q_h(W)$ is the optimal heat transferred under condition that the
consumed work is not larger than $W>0$, while $\Q_h^{\rm [sp]}$ is the
optimal spontaneous heat; see (\ref{tver}, \ref{qsp}).  Note that the
two subtracted quantities $\Q_h(W)$ and $\Q_h^{\rm [sp]}$ in
(\ref{effa}) refer to the same junction $\R+\S$, but with different
Hamiltonians; see (\ref{tver}, \ref{tver1}). 

For $W\to 0$, $\chi(W)$ increases monotonically and tends to a well
defined limit $\chi(0)$; see Fig. \ref{f2}. 
\begin{itemize}

\item For fixed $\theta$ and
$n$, $\chi(0)=\chi(W\to 0)$ is the maximal possible efficiency at which the
enhanced heat pump can operate.  As seen from Fig. \ref{f2}, this
maximum is reached for
\BEA
\Q_h(W)-\Q_h(0)\to +0\quad {\rm and}\quad W\to +0,
\label{duba}
\EEA
where we recall that $n$, $T_h$ and $T_c$ are held fixed.

\item
There is thus a complementarity between the driven 
contribution in the heat, which according to (\ref{elvonc})
maximizes for $W\to \infty$, and the efficiency that maximizes 
under $W\to 0$.

\end{itemize}

\begin{figure}
\includegraphics[width=8.5cm]{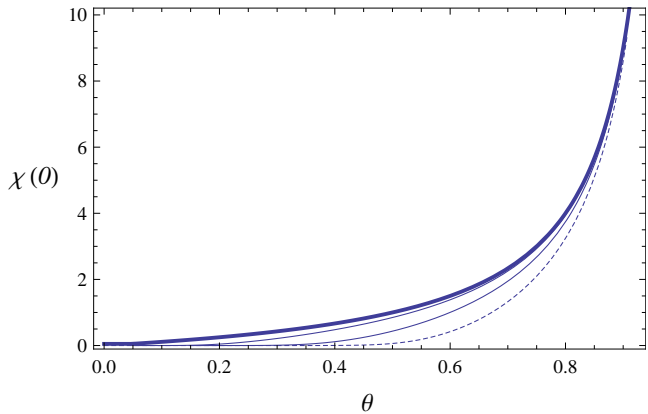}
\caption{ The maximal efficiency $\chi(0)=\chi(W=0)$ given by (\ref{kavkaz}) versus 
$\theta=T_c/T_h$ for $n=2$ (top normal curve), $n=101$ (bottom normal curve), and $n=10^5$
(dotted curve). Thick curve: the efficiency $\theta/(1-\theta)$.}
\label{f3}
\end{figure}

Note from Fig.~\ref{f3} the following aspect of the maximal
efficiency $\chi(0)$: it decreases for a larger $n$ (and a fixed
$\theta$). This is related to the fact that the optimal spontaneous heat
$\Q_h^{\rm [sp]}$ increases for larger $n$. 

\begin{itemize}

\item
It is seen from Fig.~\ref{f2} that 
\BEA
\chi(W)\leq \chi(0)<\frac{\theta}{1-\theta}.
\label{dubinushka}
\EEA
We checked that this upper bound for the efficiency (\ref{effa}) holds for all
$\theta=T_c/T_h$ and $n$.  

\end{itemize}

It will be seen below that the upper bound $\frac{\theta}{1-\theta}$ is
reached in the quasi-equilibrium limit $\theta\to 1$. Note that
$\frac{\theta}{1-\theta}$ formally coicides with the Carnot limiting
efficiency for ordinary refrigerators; see (\ref{1}). A straightforward
implication of (\ref{dubinushka}) is that enhancing optimal spontaneous
processes must be inefficient for $\theta\to 0$. 

Let us discuss to which extent the bound (\ref{dubinushka}) is similar
to the Carnot bound (\ref{1}) for refrigerators. 

{\bf 0.} These two expressions are formally identical. 

{\bf 1.} Recall that (\ref{1}) is a general upper bound for the
efficiency of refrigerators that transfer heat against its gradient.
Such a transfer does require work-consumption. The same aspect is
present in (\ref{effa}), because by its very construction the efficiency
(\ref{effa}) refers to enhancement of the optimal spontaneous process
that also demands work-consumption. To clarify this point consider a
spontaneous process with the transferred heat $Q_h^{\rm [sp]}$. Let this
spontaneous process be non-optimal in the sense that no full
optimization over the Hamiltonians (\ref{202}, \ref{co}) has been
carried out: $Q_h^{\rm [sp]}<\Q_h^{\rm [sp]}$. This non-optimal process
is now enhanced via a work-consuming one. Denote by $Q_h(W)>Q_h^{\rm
[sp]}$ the transferred heat of this process, where $W$ is the consumed
work.  Following (\ref{effa}) one can define the efficiency of this
enhancement as $\chi'(W)=[Q_h (W)-Q_h^{\rm [sp]}]/W$.  One can now show,
see Appendix \ref{etru}, that $\chi'(W)$ can be arbitrary large for a
sufficiently small (but non-zero) consumed work $W$. The reason for this
unboundness is that we consider a non-optimal spontaneous process, which
can be also enhanced by going to another spontaneous process. 

{\bf 2.} We noted above that reaching bound (\ref{dubinushka}) means a
neglegible enhancement; see (\ref{duba}). The same holds for the Carnot
bound (\ref{1}) for refrigerators: operating sharply at the Carnot
efficiency means that the heat transferred during refrigeration is zero;
see \cite{AKM} and references therein. 

{\bf 3.} An obvious point where the bounds (\ref{dubinushka}) and (\ref{1})
differ from each other is that the latter is a straightforward
implication of the first and second laws of thermodynamics, while the
former is so far obtained in a concrete model only.  We opine however
that its applicability domain is larger than this model; some support
for this opinion is discussed in section \ref{baratino}. 

\subsection{Efficiency in the quasi-equilibrium limit $\theta\to 1$.}

Let us now find a more explicit expression for $\chi(0)$ using the fact
of (\ref{duba}).  For a small $\frac{W}{T_h}$, (\ref{tver1}) implies
that $u$ and $v$ are related as $u=v^\theta+x$, with
$x={\cal O}[\frac{W}{T_h}]$. Denote the RHS of (\ref{tver}) as 
$f[u,v]$ and the RHS of (\ref{tver1}) as $g[u,v]$.
Expand $g[v^\theta+x,v]$ over $x$, express $x$ via $\frac{W}{T_h}$ and $v$:
$$x= \frac{W}{T_h}\,\frac{1}{g'_u[v^\theta,v]},  $$
and substitute the obtained result into (\ref{tver}): 
$$ f[v^\theta,v]+\frac{W}{T_h}\,\frac{f'_u[v^\theta,v] }{g'_u[v^\theta,v]}. $$
Carrying out the
optimization of this quantity over $v$, we see that in the first order over the small parameter $\frac{W}{T_h}$ the optimal value reads 
$$f[v_0^\theta,v_0]+\frac{W}{T_h}\,\frac{f'_u[v_0^\theta,v_0] }{g'_u[v_0^\theta,v_0]},  $$
where $v_0$ is defined in (\ref{qsp}).
The final result for the efficiency $\chi(0)$ is $\frac{f'_u[v_0^\theta,v_0] }{g'_u[v_0^\theta,v_0]}$, or
\BEA
\label{kavkaz}
\chi(0)=-1+\theta \ln\left(\frac{1}{v_0}\right)\left[\frac{1}{1+(n-1)v_0^\theta}-\frac{v_0}{v_0-v_0^\theta}\right].
\EEA

Let us now find the asymptotic behaviour of the efficiency (\ref{kavkaz}) for $\theta\to 1$. It
is seen from (\ref{qsp}) that in this limit $v_0=a+{\cal O}(1-\theta)$, where $a$ is found from the
transcendental equation
\BEA
\label{a}
\ln \frac{1}{a}=2\,\frac{1+(n-1)a}{1-(n-1)a}.
\EEA
This implies for $\chi(0)$ and $\Q_h^{\rm [sp]}$:
\BEA
\label{dan}
&&\chi(0)= \frac{\theta}{1-\theta}-1+\frac{2\theta}{1-(n-1)a}+{\cal O}(1-\theta),\\
&&\Q_h^{\rm [sp]}=\frac{4T_c\,(n-1)\,a\,(1-\theta)}{(1-(n-1)a)^2}+{\cal O}([1-\theta]^2).
\label{dum}
\EEA
Thus, in the limit $\theta\to 1$ the efficiency $\chi$ asymptotically
reaches the maximal value $\frac{\theta}{1-\theta}$.
This reachability is related to the fact that the
spontaneous heat $\Q_h^{\rm [sp]}$ becomes small; see (\ref{dum}). 

Note that the aymptotics (\ref{dan}) is meaningful only for $\theta\to
1$, where $\frac{\theta}{1-\theta}$ is larger than
$-1+\frac{2\theta}{1-(n-1)a}$. For $\ln n\gg 1$, (\ref{a}) implies
asymptotically
\BEA
a(n-1)=1-\frac{4}{\ln[n-1]}+\ldots.
\EEA 
Thus the limits $\theta\to 1$ and $n\gg 1$ do not commute with each other, because the
correction term $\frac{2\theta}{1-(n-1)a}$ in (\ref{dan}) will be
comparable to the main term $\frac{\theta}{1-\theta}$. 

\section{Enhanced heat transfer in linear non-equilibrium thermodynamics}
\label{baratino}

Since the above results were obtained on a concrete model, one can
naturally question their general validity.  Here we indicate that these
results are recovered from the formalism of linear non-equilibrium
thermodynamics \cite{kaplan,mazur,strat}. This theory deals with two
coupled processes: heat transfer between two thermal baths and work done
by an external field.  In contrast to the model studied in previous
sections, the field is not time-dependent; e.g., it can be associated
with the chemical potential difference \cite{strat}.  The difference and
similarity between set-ups, where the work is induced by a
time-dependent field, and those where the work is done by a constant
field is discussed in \cite{tammy}. For mesoscopic models which lead
to linear non-equilibrium thermodynamics see, e.g., 
\cite{strat,komatsu}. 

In the linear regime both the driving field and the temperature
difference $T_h-T_c$ are small, so that the heat $Q_h$ and work $W$ can
be linearized \cite{kaplan,mazur,strat}; see also Appendix
\ref{onsager}.  The virtue of this approach is that it is independent
from the concrete details of the studied system. The role of a free
parameter|over which the setup can be optimized|is played here by the
phenomenological coupling between the heat transfer and work input
\cite{kaplan}. 

More specifically, we show in Appendix \ref{onsager} that also within
the formalism of linear non-equilibrium thermodynamics, enhancing the
optimal spontaneous flow requires work-consumption. As for the
efficiency, we first recall the message of (\ref{dan}): for $\theta\equiv
T_c/T_h\to 1$ the maximal efficiency $\chi(0)$ does not depend on the
dimension $n^2$ of the junction (provided that the latter is fixed) and
approaches $\propto 1/(1-\theta)$. This asymptotic result is to a
certain extent universal, because it is also recovered within linear
non-equilibrium thermodynamics; see Appendix \ref{onsager}. We stress
however that only the asymptotics for $\theta\to 1$ is recovered. Since
this is a linear theory restricted to a small temperature difference and
a small work input, it is naturally not capable of reproducing the full
message of the bound (\ref{dubinushka}). For this purpose one would
probably need a non-linear thermodynamics theory; see, e.g.,
\cite{strat}.  Unfortunately, such theories are not so universal
(system-independent) as the linear theory.

\section{Summary}
\label{conc}

We started this paper by listing several representative examples of
enhanced transport and posing two basic questions:

\begin{enumerate}

\item When enhancing a spontaneous process does require
work-consumption? 

\item If such a work-driven enhancement does take place, is there a
general bound on its efficiency? 

\end{enumerate}

These questions have been answered via a quantum model for a
heat-transfer junction immersed between two thermal baths at different
temperatures $T_c$ and $T_h$ ($T_c<T_h$). The model is defined in
section \ref{model}. We have chosen to work with this model
of junction, because its structure is flexible enough to allow explicit
maximization of the transferred heat over the junction Hamiltonian.  We
thus can determine the maximal heat transferred by the junction.

Our basic results can be stated as follows.

\begin{enumerate}

\item When the spontaneous heat-transfer is already maximized over
the junction Hamiltonian, its enhancement does require work-consumption. 

\item The efficiency is defined with respect to the optimal spontaneous
heat transfer as the heat increment due to ehancement divided over the
consumed work. This efficiency is shown to be limited from above by
$T_c/(T_h-T_c)$, a bound that is reached for $T_c\to T_h$. For this
bound to hold it is essential that the efficiency is defined with
respect to optimal spontaneous heat transfer. In its turn, the very idea
of optimality refers to a class of variables to be optimized
over. For the studied model this class is defined by the junctions
structure and its Hamiltonian. 

\end{enumerate}

The main open problem with these results is whether they hold more
generally than the studied model. We presented a partial
evidence|partial because it is restricted to a small $T_h-T_c$ within
the linear non-equilibrium thermodynamics|that they hold more
generally.

\subsection*{Acknowledgement}
This work has been supported by Volkswagenstiftung.

We thank B. Mehmani for several useful suggestions.

\appendix

\section{Power of heat transfer}
\label{po}

Recall that the power $Q_h/\tau$ is defined as the
ratio of the transferred heat $Q_h$ to the cycle duration $\tau$. For
our model $\tau$ is mainly the duration of the second stage, i.e.,
$\tau$ is the relaxation time, which depends on the concrete physics of
the system-bath coupling.

We discuss two scenarios of relaxation for the present model. For
the collisional relaxation the target system interacts with
independent bath particles via successive collisions; see
\cite{partovi,mityugov,AKM}. For our purposes the target system is $\R$ or $\S$ that interact with,
respectively, the hot and cold bath. Each collision lasts a time
$\tau_{\rm col}$, which is much smaller than the characteristic time
$\tau_{\rm btw}$ between two collisions. The interaction Hamiltonian
[between the target system and a bath particle] is conserved, so that
there is no energy cost in switching the system-bath interaction on and
off; see \cite{partovi,mityugov,AKM}. The
relaxation time $\tau$ can amount to a few $\tau_{\rm btw}$: $\tau_{\rm
btw}\lesssim \tau$; see \cite{AKM}.

If now the pulse-time $\delta$ of $V(t)$ is also much smaller than
$\tau_{\rm btw}$, one realizes a thermally-isolated pulse (since the
overlap between the pulse and a collision is negligible), while the
relaxation time $\tau$ can be still much smaller than any characteristic
time of $\S$ or $\R$. Thus in this scenario the cycle time $\tau$ is
finite, and the power of heat pumping $Q_c/\tau$ does not vanish due to
a large cycle time.

Within the second scenario the system-bath interaction is always on, but
its magnitude is small [weak-coupling]. Now the relaxation time is much
larger than the internal characteristic time of $\R$ and $\S$.  Because
the system-bath interaction is always on, there will be a contribution
in the work (\ref{work}) coming from the system-bath interaction
Hamiltonian. This contribution arises even when the conditions for the
pulsed regime hold \cite{armen}. However, within the weak coupling
assumption this additional contribution is proportional to the square of
the system-bath interaction constant and can neglected \cite{armen}.  We
stress that this additional contribution does not arise within the
collisional relaxation scenario, because the pulse and collisions are
well-separated in time.

\section{}
\label{etru}

Let the system ${\bf H+C}$ functions as a {\it non-optimal} [with respect to
the Hamiltonian of $\R+\S$] spontaneous heat pump.  The amount of heat
transferred per cycle is $Q_{h}^{\rm [sp]}$. One increases the
transferred heat via an external work-source acting on ${\bf H+C}$. Now
it is equal to $Q_h>Q_{h}^{\rm [sp]}$. The amount of work consumed per
cycle is $W$. 

The purpose of this Appendix is to show on a concrete example 
that there is no upper bound on the efficiency:
\BEA
\label{pl}
\chi'=\frac{Q_h-Q_h^{\rm [sp]}}{W}.
\EEA
It can go to $+\infty$ for $W\to +0$.

Let the systems $\R$ and $\S$ be three-level systems with Hamiltonians
[compare with (\ref{2})]
\BEA
\label{kl}
H_\R={\rm diag}(0,\eps,\eps+\mu),\quad H_\S={\rm diag}(0,\mu,\eps+\mu)
\EEA
where for future purposes we choose $\mu>\eps>0$. The initial Hamiltonian (\ref{222}) reads
\BEA
\label{krot}
\diag[0,\,\eps,\,\eps+\mu,\,\mu,\,\eps+\mu,\,\eps+2\mu,\,\eps+\mu,\,2\eps+\mu,\,2\eps+2\mu]\nonumber 
\EEA

The initial state of $\R+\S$ is given by (\ref{1}, \ref{2} \ref{111}):
\BEA
\rho=\frac{\diag [1,\,\,a^\theta,\,\,a^\theta b^\theta]}{1+a^\theta+a^\theta b^\theta}
, ~~~
\sigma=\frac{\diag [1,\,\,b,\,\,ab]}{1+b+ab},
\EEA
where we denoted
\BEA
a=e^{-\beta_c \eps},\quad b=e^{-\beta_c \mu}.
\EEA

The system ${\bf H+C}$ has nine energy levels; 
three of them have equal energy $\eps+\mu$. Thus spontaneous processes 
amount to unitary operations that couple these degenerate energy levels to each other, but do not
induce transitions to non-degenerate energy levels [compare with (\ref{co})]. 
Such unitaries will produce double-stochastic matrices of the following form
[compare with (\ref{gomel}, \ref{chernigov})]
\begin{gather}
C=\lp\begin{array}{ccccccccc}
1 & 0 & 0 & 0 & 0 & 0 & 0 & 0 & 0 \\
0 & 1 & 0 & 0 & 0 & 0 & 0 & 0 & 0 \\
0 & 0 & c_{11} & 0 & c_{12} & 0 & c_{13} & 0 & 0 \\
0 & 0 & 0 & 1 & 0 & 0 & 0 & 0 & 0 \\
0 & 0 & c_{21} & 0 & c_{22} & 0 & c_{23} & 0 & 0 \\
0 & 0 & 0 & 0 & 0 & 1 & 0 & 0 & 0 \\
0 & 0 & c_{31} & 0 & c_{32} & 0 & c_{33} & 0 & 0 \\
0 & 0 & 0 & 0 & 0 & 0 & 0 & 1 & 0 \\
0 & 0 & 0 & 0 & 0 & 0 & 0 & 0 & 1
\end{array}\rp,
\end{gather}

It is not difficult to see that for {\it fixed energy levels} (\ref{kl})
the spontanoues process with the largest $Q_h^{\rm [sp]}$ amounts to
$c_{13}=c_{22}=c_{31}=1$, while other $c$'s nullify. This means that the
spontaneous process interchanges populations of the third and seventh
energy levels; see (\ref{krot}). The heat transferred per cycle reads
\BEA
Q_{h}^{\rm [sp]}=\frac{1}{z}{\lb a^\theta b^\theta - a b\rb(\eps + \mu)},\\
z\equiv {\lp 1+a+a b\rp\lp 1+a^\theta + a^\theta b^\theta\rp}.
\EEA

Now the external field acts on the system ${\bf H+C}$ enhancing the
heat transfer, i.e., increasing $Q_{h}^{\rm [sp]}$. We postulate that
this action amounts to SWAP transformation.  We obtain for the heat,
work and efficiency (\ref{pl}) [compare with (\ref{heats}, \ref{work})]:
\begin{gather} \nonumber
Q_h = \frac{1}{z} \{ \eps \lb a^\theta-b+ba^{\theta}(a-b^\theta) \rb 
\\
+ (\eps+\mu) \lb (a b)^\theta - a b + ba^{\theta} (b^\theta - a) \rb \},
\nonumber\\
W = \frac{1}{z}(\mu-\eps)\lb a^\theta-b-ba^{\theta}(b^\theta-a) \rb
\nonumber
\end{gather}
It is seen that when $\eps\to\mu$ [which means $a\to b+0$] the
difference $Q_h-Q_h^{\rm [sp]}$ is positive and finite, while the work
$W$ is positive, but turns to zero as $W\propto \mu-\eps$. Thus if one
defines efficiency as $\chi'=(Q_h-Q_h^{\rm [sp]})/W$, it will turn to
infinity for $\eps\to\mu$. 
The divergence of $\chi'$ is ultimately due to the fact that the
spontaneous heat was maximized over Hamiltonian (\ref{202}) only
partially, i.e., it was not maximized over the energy levels of $\R$ and
$\S$. 

\section{Heat-transfer enhancement from linear non-equilibrium thermodynamics}
\label{onsager}

The formalism of linear non-equilibrium thermodynamics starts by
introducing currents of physical quantities $J_i$ (e.g., currents of
energy or mass) and the respective conjugate forces $X_i$
\cite{kaplan,mazur}. We introduce two such currents and forces:
\BEA
J_1,\,\, J_2,\,\, X_1,\,\, X_2.
\EEA
$J_1$ will refer to heat flowing from the
high-temperature bath at temperature $T_h=1/\beta_h$ to the lower one at
temperature $T_c=1/\beta_c$, while the second current $J_2$ is the 
work done by an external time-independent force. Thus
\BEA
\label{ba}
X_1=\beta_c-\beta_h=\frac{T_h-T_c}{T^2}, ~~ X_2=\frac{f}{T}, ~~ 
T\equiv \sqrt{T_hT_c},
\EEA
where $f$ is the
force responsible for the work \cite{mazur}. Once we are in the
linear regime over small parameters $T_h-T_c$ and $f$, one can substitute $T$ in (\ref{ba}) by 
$T_c$ or $T_h$; the choice of $T$ is conventional.

In essence, linearity means that the state which supports the currents
is not far from equilibrium \cite{mazur}. The basic postulate of this
formalism is linear relations between currents and forces \cite{mazur}:
\BEA
J_1 = L_{11}X_1 + L_{12}X_2, \qquad J_2 = L_{21}X_1 + L_{22}X_2.
\EEA
where the kinetic coeffcients $L_{ik}$ do not depend on 
$X_i$ due to assumed linearity of the overall process.  
The kinetic coeffcients $L_{12}$ and $L_{21}$ quantify the coupling between the two processes.

The statement of the second law relevant for this composite linear process amounts to
the positivity of entropy production $\d_iS/\d t$ \cite{mazur}
\BEA
\label{o3}
\frac{\d_iS}{\d t}= X_1J_1 + X_2J_2=\sum_{i,k=1}^2 L_{ik}X_iX_k\geq 0.
\EEA
As a consequence of the time-invariance of
the underlying microscopic theories (i.e., classical or quantum mechanics), the matrix of kinetic
coefficients is symmetric \cite{mazur}
\BEA
\label{o4}
L_{12} = L_{21}.
\EEA

We now assume that some work is dissipated, $J_2>0$, for enhancing the
heat flow. The spontaneous heat transfer corresponds to no coupling between the processes: $L_{12} = L_{21}=0$.

In calculating the efficiency of enhancement we shall follow the same
strategy as in sections \ref{QforE} and \ref{effi}: first we shall
maximize the transferred heat $J_1$ under a fixed amount of work $W=fJ_2=
TX_2 J_2$.  Then the efficiency will be defined as in (\ref{effa}). The
maximization variables are $f$, $L_{11}$, $L_{22}$ and $L_{12}=L_{21}$.
The temperatures $T_c$ and $T_h$ are held fixed; see also
(\ref{ba}). 
Write $J_1$ as
\BEA
J_1= L_{11}X_1+\frac{W}{TX_1}-L_{22}\frac{X_2^2}{X_1}.
\EEA
During the maximization we should keep $L_{11}$ confined by some upper limit ${\cal L}_{11}$; 
otherwise $J_1$ will not be finite. Eq.~(\ref{o4}), which should hold for arbitrary $X_1$ and $X_2$, implies
$L_{22}\geq 0$. Recaling that also $X_1>0$, we see that $J_1$ is maximized for $L_{11}={\cal L}_{11}$
and $L_{22}=0$:
\BEA
{\cal J}_1= {\cal L}_{11}X_1+\frac{W}{TX_1}.
\EEA
Thus for enhancing the optimal spontaneous heat ${\cal L}_{11}X_1$ we need $W>0$ (work-consumption).
Subtracting from ${\cal J}_1$ the spontaneous contribution ${\cal L}_{11}X_1$ and dividing over the consumed work $W$,
we get for the efficiency:
\BEA
\chi=\frac{1}{TX_1}=\frac{T}{T_h-T_c},
\EEA
which for a small temperature difference $T_h-T_c$ (recall that this is
the applicability domain of the considered linear theory) coincides with (\ref{dan}).

\end{document}